\documentclass[apj]{emulateapj}

\newcommand {\Lya}    {Ly$\alpha$}   
\newcommand {\Lyb}    {Ly$\beta$}    
\newcommand {\OVI}    {\ion{O}{6}}
\newcommand {\OVII}   {\ion{O}{7}}
\newcommand {\OVIII}  {\ion{O}{8}}
\newcommand {\NV}     {\ion{N}{5}}

\newcommand {\HI}     {\ion{H}{1}}   

\newcommand {\SiIII}  {\ion{Si}{3}}
\newcommand {\CIV}    {\ion{C}{4}}

\newcommand {\kms}    {km~s$^{-1}$}
\newcommand {\NHI}    {$N_{\rm HI}$}

\newcommand {\lam}    {$\lambda$}
\newcommand {\FUSE}   {{\it FUSE}} 
\newcommand {\HST}    {{\it HST}}
\newcommand {\Chandra}{{\it Chandra}}
\newcommand {\XMM}    {{\it XMM-Newton}}
\newcommand {\etal}   {et~al.} 
\newcommand {\cd}     {cm$^{-2}$}  
\newcommand {\flux}   {$\rm erg~cm^{-2}~s^{-1}~\AA^{-1}$}
\newcommand {\tnma}   {\tablenotemark{a}}
\newcommand {\tnmb}   {\tablenotemark{b}}
\newcommand {\tnmc}   {\tablenotemark{c}}
\newcommand {\tnmd}   {\tablenotemark{d}}

\slugcomment{ApJ accepted July 2011}

\begin{document}

\title{An HST/COS Search for Warm-Hot Baryons in the Mrk\,421 Sightline
\footnote{Based on observations made with the NASA/ESA {\it Hubble Space Telescope}, obtained from the data archive at the Space Telescope Science Institute.  STScI is operated by the Association of Universities for Research in Astronomy, Inc. under NASA contract NAS5-26555.} 
}

\author{Charles W. Danforth, John T. Stocke, Brian A. Keeney, Steven V. Penton, J. Michael Shull, Yangsen Yao \& James C. Green}
\affil{CASA, Department of Astrophysical and Planetary Sciences, University of Colorado, 389-UCB, Boulder, CO 80309; danforth@colorado.edu}

\begin{abstract}

Thermally-broadened \Lya\ absorbers (BLAs) offer an alternate method to using highly-ionized metal absorbers (\OVI, \OVII, etc.) to probe the warm-hot intergalactic medium (WHIM, $T=10^5-10^7$~K).  Until now, WHIM surveys via BLAs have been no less ambiguous than those via far-UV and X-ray metal-ion probes.  Detecting these weak, broad features requires background sources with a well-characterized far-UV continuum and data of very high quality.  However, a recent HST/COS observation of the $z=0.03$ blazar Mrk\,421 allows us to perform a metal-independent search for WHIM gas with unprecedented precision.  The data have high signal-to-noise ($S/N\approx50$ per $\sim20$~\kms\ resolution element) and the smooth, power-law blazar spectrum allows a fully-parametric continuum model.  We analyze the Mrk\,421 sight line for BLA absorbers, particularly for counterparts to the proposed \OVII\ WHIM systems reported by \citet{Nicastro05a,Nicastro05b} based on \Chandra/LETG observations.  We derive the \Lya\ profiles predicted by the X-ray observations.  The signal-to-noise ratio of the COS data is high ($S/N\approx25$ per pixel), but much higher $S/N$ can be obtained by binning the data to widths characteristic of the expected BLA profiles.  With this technique, we are sensitive to WHIM gas over a large ($N_H,T$) parameter range in the Mrk\,421 sight line.  We rule out the claimed \citeauthor{Nicastro05a} \OVII\ detections at their nominal temperatures ($T\sim1-2\times10^6$~K) and metallicities ($Z=0.1\,Z_\sun$) at $\ga2\sigma$ level.  However, WHIM gas at higher temperatures and/or higher metallicities is consistent with our COS non-detections.
\end{abstract}



\section{Introduction}

The Warm-Hot Intergalactic Medium (WHIM) is expected to make up 25--50\% of the baryons in the low-redshift Universe based on cosmological simulations \citep{CenOstriker99,Dave99,Dave01,Smith11}.  Thus far, only a small fraction of this hot gas has been detected \citep[see, e.g.][]{Danforth09}, and the interpretation of these detections is still controversial.  The difficulty in detecting this gas lies in its combination of high temperature ($10^{5-7}$~K) and low hydrogen density ($10^{-4}-10^{-6}\rm~cm^{-3}$).  For example, only the subset of WHIM gas which is sufficiently metal-enriched can be detected through highly-ionized metal ions at UV and X-ray energies.   The \OVI\ \lam\lam1031, 1037 doublet has been detected in numerous extragalactic systems with the {\it Far Ultraviolet Spectroscopic Explorer} (\FUSE) and {\it Hubble Space Telescope} (\HST) \citep{DS05,DS08,Tripp08,ThomChen08}.  It may trace very low density, photoionized $\sim10^4$~K gas \citep[e.g.,][]{OppenheimerDave08,Oppenheimer11} or shock-heated, collisionally-ionized WHIM \citep{Smith11}.  Until now, the searches for hotter WHIM ($>10^6$~K) have been restricted to soft X-ray absorption. 
 
Very highly ionized metal species such as \OVII, \OVIII, \ion{N}{7} and \ion{Ne}{9} with soft-X-ray resonance transitions offer a good option for finding diffuse gas at the hotter end of the WHIM temperature range.  However, the predicted low metallicity of this gas \citep[$Z\sim 0.1~Z_\sun$ typical of \OVI\ absorbers;][]{DS08} and the relative insensitivity and low resolution of even our best, current X-ray spectrometers (i.e., the \Chandra\ Low Energy Transmission Grating (LETG) and \XMM\ Reflection Grating Spectrometer (RGS)) has generated controversy for most or all current proposed detections of hotter WHIM gas.  

\begin{figure*}
  \epsscale{1}\plotone{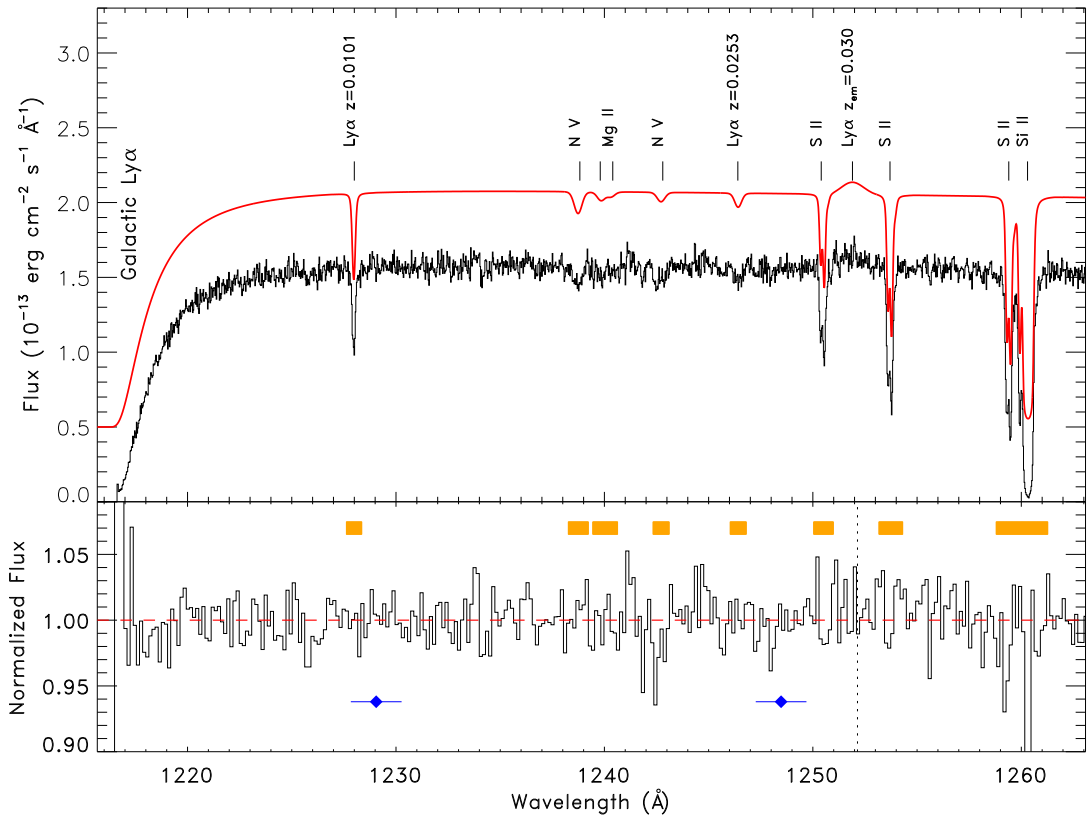} 
  \caption{HST/COS spectrum of Mrk\,421 spanning the entire redshift
  pathlength of the \Lya\ forest ($1216 \AA<\lambda<1252$ \AA).  The
  data in the top panel are binned by three pixels ($\sim1/2$
  resolution element).  The power-law continuum, Galactic damped \Lya\
  absorption, intrinsic \Lya\ emission ($z_{em}=0.030$), and known
  narrow absorption features detailed in the text are modeled by the
  red curve offset from the flux by $+5\times10^{-14}$ \flux.  The
  lower panel shows the residual data binned by seven pixels
  (approximately the COS resolution element) after the continuum and
  all absorption/emission features are divided out.  The regions where
  narrow lines have been modeled out of the data are shown by orange
  bars.  Blue diamonds mark the redshift locations and $\pm300$ \kms\
  uncertainties of the \citet{Nicastro05a} claimed O\,VII features.}
\end{figure*}

The growing number of ``broad \Lya\ absorbers''(BLAs) detected in high signal-to-noise FUV spectra \citep{Sembach04,Richter04,Richter06a,Lehner07,Danforth10a} may trace WHIM independent of metal enrichment, but this diagnostic suffers from a different set of observational uncertainties.  Even at the cooler end of the WHIM temperature range, the fraction of hydrogen in a neutral state is tiny.  Thermal broadening renders any \Lya\ absorption shallow and broad, and the WHIM absorber may be blended with narrower absorption features arising in adjacent, cooler, photoionized gas.  At temperatures above $10^6$~K, spectra of extremely high quality, with well-defined continua, are required to characterize the WHIM through BLAs alone.  Independent measurements of individual WHIM systems through complimentary techniques reduce the uncertainties inherent in single-method detections.
  
Two of the earliest and most-cited X-ray WHIM detections were the two reported \OVII\ absorbers \citep{Nicastro05a,Nicastro05b} in \Chandra\ LETG observations of the blazar Mrk\,421 ($z_{em}=0.030)$.  These observations were made while Mrk\,421 was in a flaring state and represent some of the highest signal-to-noise soft-X-ray data taken to date on an extragalactic sight line.  Although these absorbers were {\it not confirmed} in observations by \XMM\ \citep{Kaastra06,Williams06,Rasmussen07}, their presence has neither been confirmed nor ruled out conclusively.  While a few other X-ray WHIM absorption detections have been proposed \citep[e.g., in the PKS\,2155-304 sight line by][]{Fang02}, a new tactic employed by \citet{Buote09} may have finally achieved some success.  Based on the results of simulations \citep[e.g.,][]{Dave99} that show high concentrations of WHIM gas around large-scale filamentary structure, \citet{Buote09} \citep[see also][]{Fang10,Zappacosta10} observed the bright blazar H\,2356$-$309 behind the Sculptor Wall and made \OVII\ detections at the redshift of that filament with both \Chandra\ and \XMM.  Based on the Sculptor Wall detection, it seems worthwhile to revisit the Mrk\,421 sight line, since one of the two claimed WHIM detections is at the redshift ($z=0.027$) of another nearby, large-scale galaxy filament, the Great Wall \citep{deLapparent86,Penton00,Williams10}.
  
In this paper, we focus on the potential for detecting WHIM in high-quality UV spectra from the Cosmic Origins Spectrograph (COS) on HST.  This analysis is specific to the Mrk\,421 sight line, but easily generalized to sight lines for which high-quality far-UV spectra exist.  The FUV spectra of blazars can be well-modeled as a power law \citep{Stocke11}, and the high $S/N$ and multiple ionic transitions allow foreground Galactic absorbers to be well-modeled.  Since the data can be fitted with a fully parametric continuum model, this is an ideal test-bed for detecting WHIM gas in BLAs.  

In Section 2, we present the COS FUV spectrum of Mrk\,421, identify all spectral features present, both Galactic and extragalactic, and produce a normalized, line-free spectrum covering the entire \Lya\ forest region.  In Section~3, we model the BLA absorbers implied by both the reported \citet{Nicastro05a,Nicastro05b} (henceforth collectively N05) \OVII\ measurements and for more generalized WHIM absorbers.  We then place limits on the temperature and metallicity of any WHIM gas present along the Mrk\,421 line of sight based on the observed COS data.  In Section~4 we discuss the meaning and importance of these results, reiterate our main conclusions, and discuss applications to other sight-lines.  


\section{Observations and Analysis}

COS far-UV observations of BL\,Lac object Mrk\,421 ($z_{em}=0.030$) were carried out in 2009, December, during the first three months of COS science observations as part of the COS Guaranteed Time Observations (PID 11520, PI Green).  Four exposures were made in the G130M (1135 \AA\ $<\lambda<$ 1480~\AA; totalling 1.7 ksec) medium-resolution grating ($R\approx18,000$), each at a different central wavelength setting, to dither over known instrumental features \citep[see][]{Green11,Osterman11}. 

The exposures were reduced with a custom flat field derived from the combination of several public COS calibration observations of white dwarfs.  Four targets were used to create the G130M flat field; WD\,0308$-$565, WD\,0320$-$539, WD\,0947$+$857, and WD\,1057$+$719.  For each calibration target, an iterative technique was used to independently measure the relatively smooth continua from the inferred underlying flat-field and grid wire shadows using the `x1d' files output by {\sc CALCOS}~2.13.  The four independent one-dimensional estimates of the G130M flat field were merged in detector space with $(S/N)^2$ weighting.  After applying the custom flat field to each exposure, we aligned the four Mrk\,421 exposures in wavelength space via a cross-correlation of regions around prominent ISM absorption features.  Exposures were interpolated onto a common wavelength grid and combined via an exposure time weighted mean as described in \citet{Danforth10b}.  

Our custom flat field is not perfect, but it consistently improved the continuum signal-to-noise in our Mrk\,421 spectrum as compared to the standard ``wireflat'' processing \citep{Danforth10b}.  For example, in the 1410--1430 \AA\ region of the Mrk\,421 G130M spectrum, where no ISM or IGM absorption lines are expected, the signal-to-noise per seven-pixel resolution element improved from 27 to 36 with our custom flat fielding, as compared to our standard ``wireflat'' processing.  A second line-free region from 1340--1369 \AA\ increased from S/N of 38 to 48 per resolution element. 

\subsection{Spectral Modeling}

The \Lya\ forest region of the Mrk\,421 sight-line is shown in the top panel of Figure~1.  Since we are interested in weak, broad absorption features in the data, an accurate and detailed model fit to the spectrum including all known features is crucial.  Fortunately, as with most blazars, the dereddened UV continuum is easily defined by a power law of the form $F_\lambda=F_{912}\,(\lambda/912\AA)^{-\alpha_\lambda}$.  We use the parameters $F_0=3.405\times10^{-13}$ \flux\ and $\alpha_\lambda=1.918\pm0.020$ to provide a baseline continuum fit to the data in the 1150 \AA--1300 \AA\ region.  Note that these parameters are slightly different from those reported by \citet{Stocke11}, which were based on an earlier reduction of the data and a continuum fit to the entire COS/FUV spectral region.  The damped Galactic \Lya\ absorber is modeled with the parameters $v=-28\pm1$ \kms, $b_{\rm HI}=55\pm2$ \kms, and $\log\,N\rm_{HI}\,(cm^{-2})=20.042\pm0.002$.  This is consistent with the emission-weighted mean \HI\ 21 cm profile in a $\sim0.6^\circ$ beam toward Mrk\,421 in the Leiden-Argentine-Bonn \HI\ survey \citep[LAB,][]{LABsurvey}.  The measured column of neutral hydrogen corresponds to a Galactic reddening of $E(B-V)=N_{\rm HI}/5.8\times10^{21}~\rm cm^{-2}=0.019$.  (A fit to the Galactic \Lyb\ profile in the \FUSE\ data yields a consistent value.)  In a departure from the pure power-law source continuum typical of BL\,Lac objects, Mrk\,421 shows a weak emission feature at 1251.9~\AA\ corresponding to \Lya\ emission at $z_{em}=0.0298$ from circumnuclear gas in the blazar.  We include the fit from \citet{Stocke11} ($FWHM=1.23$ \AA, $I=1.27\times10^{-14}\rm~erg~cm^{-2}~s^{-1}$) in our spectral model, shown offset from the data in the top panel of Figure~1.

There are several narrow Galactic absorption lines which can be modeled out of the spectrum.  The stronger blue line of the \ion{N}{5} $\lambda\lambda 1238.82, 1242.80$ doublet is well fitted with parameters $v=-21\pm6$ \kms, $b=59\pm8$ \kms, and $\log\,N_{\rm NV}=13.34\pm0.05$, but the weaker red line of the doublet shows an absorption excess that is discussed below.  The \ion{S}{2} $1259.52,1253.81,1250.58$ \AA\ triplet has a strong Galactic component ($v=-8\pm2$ \kms, $b=13\pm4$ \kms, $\log\,N_{\rm SII}=15.05\pm0.10$), a weaker, blue-shifted component ($v=-48\pm2$ \kms, $b=14\pm5$ \kms, $\log\,N_{\rm SII}=14.89\pm0.05$), and very weak red wing ($v=32$ \kms, $b=20$ \kms, $\log\,N_{\rm SII}=13.9\pm0.1$).  These low-ion velocity components are consistent with the two \HI\ emission components \citep{Wakker03,LABsurvey} and absorption seen in other low-ionization species in COS data \citep{Yao11a}.  The strong \ion{Si}{2} $\lambda1260.42$ line, while redward of the \Lya\ forest region in Mrk\,421, can be fitted with two components ($v=-21\pm3$ \kms, $b=28\pm3$ \kms, $\log\,N_{\rm SiII}=15.2\pm0.1$) and ($v=-117\pm5$ \kms, $b=14\pm5$ \kms, $\log\,N_{\rm SiII}=13.1\pm0.1$).  The last identified Galactic absorber is the weak, blended \ion{Mg}{2} $\lambda\lambda1239.93,1240.40$ doublet.  Assuming a 2:1 ratio of equivalent widths and similar velocity centroids and $b$-values to the other Galactic lines, we model the system with $v=-21$ \kms, $b=59$ \kms, and $\log\,N_{\rm MgII}=15.3$.  

The sight line toward Mrk\,421 is relatively short ($\Delta z=0.030$), and there is only a single previously-reported \citep{Shull96} intervening \Lya\ forest system ($z=0.01014$, $\lambda=1228.0$~\AA).  This moderate-strength \Lya\ line is well-fitted by a Voigt profile with parameters ($z=0.01013$, $b=19\pm1$~\kms, $W_\lambda=76\pm4$~m\AA, $\log\,N_{\rm HI}=13.25\pm0.02$) convolved with the COS line spread function \citep{Ghavamian09,Kriss11}.  No corresponding metal absorption (e.g., \CIV, \SiIII, etc.) is seen at this redshift.

A weak absorption feature ($W_\lambda=25\pm4$ m\AA) is seen at 1246.4 \AA\ in the coadded spectrum.  There are no species commonly seen in the ISM with absorption lines near 1246 \AA, and the normalized spectrum in the $1243-1250$ \AA\ range is free of line-removal artifacts.  The most likely identification of this feature is a weak \Lya\ forest system with fit parameters $cz=7582\pm8$ \kms\ ($z=0.0253$), $b=52\pm12$ \kms, and $\log\,N_{\rm HI}=12.7\pm0.1$.

The spectrum is divided by the model (offset curve in the top panel of Figure~1) to give a normalized spectrum, with all identified absorption and emission features removed (Figure~1, lower panel).  With the exception of a few narrow line-removal artifacts, the normalized spectrum shows a remarkably uniform appearance over the range 1218~\AA\ $\la\lambda\la$ 1257~\AA.

\begin{deluxetable}{lllll}
  \tabletypesize{\footnotesize}
  \tablecolumns{5} 
  \tablewidth{0pt} 
  \tablecaption{Mrk\,421 WHIM Absorber Parameters}
  \tablehead{
           \colhead{Quantity}  		&
           \colhead{$z=0.011$}          &
	   \colhead{$z=0.027$}          & 
	   \colhead{Units} 		&
           \colhead{Source\tnma}		
            }
\startdata 
\cutinhead{UV/X-ray Measurements}
$cz$                  & $ 3300\pm300 $  &$ 8090\pm300$ & \kms          & N05 \\
$N\rm_{O\,VII}$       & $ 1.0\pm0.3  $  &$ 0.7\pm0.3 $ & $10^{15}$ \cd & N05 \\
$N\rm_{O\,VI}$        & $ <1.5       $  &$ <1.5      $ & $10^{13}$ \cd & S05 \\
$N\rm_{N\,VII}$       & $ 0.8\pm0.4  $  &$ 1.4\pm0.5 $ & $10^{15}$ \cd & N05 \\
$\log\,T$             & $ 6.13\pm0.39$\tnmb&$ 6.15\pm-0.15$\tnmc & K   & N05 \\
\cutinhead{Implied \Lya\ Absorber Parameters}
$b\rm_{therm}$        & $148^{+84}_{-53}$&$152^{+28}_{-24}$ & \kms     & \\
$\log\,N\rm_{HI}$\tnmd& $12.5^{+0.6}_{-0.2}$&$12.7\pm0.2 $  & $Z_{0.1}^{-1}$ \cd & \\  
$W\rm_{Ly\alpha}$\tnmd& $18^{+55}_{-8}$  & $ 26^{+11}_{-8}$ & $Z_{0.1}^{-1}$ m\AA & \\
$\tau_0$              & 0.016            & 0.025            &                     &  \\
$SL$                  & 1.9              & 2.7              & $\sigma$            &  \\
\enddata
\tablenotetext{a}{N05 = Nicastro \etal\ 2005a (\Chandra); S05 = Savage \etal\ 2005 (\FUSE)}
\tablenotetext{b}{Inferred from O\,VI/VII/VIII ratios}
\tablenotetext{c}{Inferred from N\,VI/VII ratios}
\tablenotetext{d}{Assuming $Z=0.1\,Z_\sun$.  See text for details.}
\end{deluxetable}

\section{WHIM Absorption Toward Mrk\,421}

Given a few input parameters, we can model the expected appearance of a BLA.  These parameters can be assumed for WHIM absorbers in general, or they can be derived from X-ray WHIM measurements and a few basic assumptions.  For general single-phase WHIM gas with total hydrogen column density $N_H$ at temperature $T$, the neutral hydrogen column density is $N_{\rm HI}=f_{\rm HI}(T)\,N_H$ and the thermal line width is 
\begin{equation}
	b(T)=(40.6~{\rm km~s^{-1}})\,T_5^{1/2},
\end{equation} 
where $T_5$ is temperature in units of $10^5$~K.  The neutral fraction as a function of temperature, $f_{\rm HI}(T)$, can be determined either in collisional ionization equilibrium (CIE) or with a more sophisticated sets of assumptions.  In the following analysis, we assume $Z=0.1\,Z_\sun$ gas cooling isobarically as tabulated in \citet{GnatSternberg07}.  At WHIM temperatures, the hydrogen neutral fraction is identical to that in CIE and the high-ion fractions differ only slightly from their CIE values.

If we start from the reported WHIM parameters derived from the \Chandra\ observations, we can derive the \NHI\ via the additional relationship,
\begin{equation}
  N_{\rm HI}=\frac{f_{\rm HI}(T)}{f_{\rm OVII}(T)} \frac{N_{\rm OVII}}{(O/H)_\sun Z}\;.
\end{equation}
Taking the reported \citeauthor{Nicastro05a} X-ray detections at face value, the $cz=3300$ \kms\ X-ray absorber is assumed to have $T$ ranging from $5.5\times10^5$~K to $3.3\times10^6$~K, based on \OVI/\OVII/\OVIII\ column density detections and limits.  The $cz=8090$ \kms\ system is inferred to have a somewhat narrower temperature range, $T=1-2\times10^6$~K, from \ion{N}{6}/\ion{N}{7} ratios and an \OVI\ non-detection \citep{Savage05}.  These WHIM parameters predict \Lya\ absorbers with column density $\log\,N_{\rm HI}=12.4-13.3$, $b=95-233$~\kms, and $W_\lambda=13-100$~m\AA\ at the $z=0.011$ absorber; and $\log\,N_{\rm HI}\approx12.7$, $b=130-180$ \kms, and $W_\lambda=10-34$~m\AA\ at $z=0.027$ (Table~1).  We assume $(O/H)_\sun=4.90\times10^{-4}$ \citep{Asplund09} and $Z=0.1\,Z_\sun$. Lower abundances of oxygen would scale these predictions upwards.

\begin{figure}
  \epsscale{1.2}\plotone{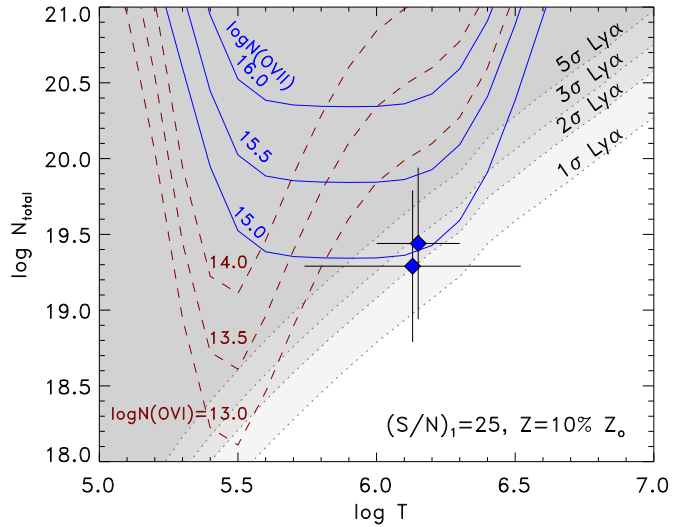}
  \caption{The detection limits for WHIM gas via broad \Lya\
  absorption depend on the parameters of the absorber and the quality
  of the far-UV spectrum.  In HST/COS data with $(S/N)_1\approx25$ per
  pixel, idealized, unblended BLAs can be detected at $1-5~\sigma$
  significance levels as shown by the dotted contours.  Typical
  detection limits for several different column densities of O\,VII
  (blue, solid) and O\,VI (red, dashed) are shown for an assumed
  metallicity of $Z=0.1\,Z_\sun$.  Metal line diagnostics scale as
  $Z^{-1}$ on the vertical axis.  Blue diamonds show the
  parameter-space location of the two \citet{Nicastro05a} claimed WHIM
  detections with their uncertainties.}
\end{figure}

\begin{figure*}
  \epsscale{.9}\plotone{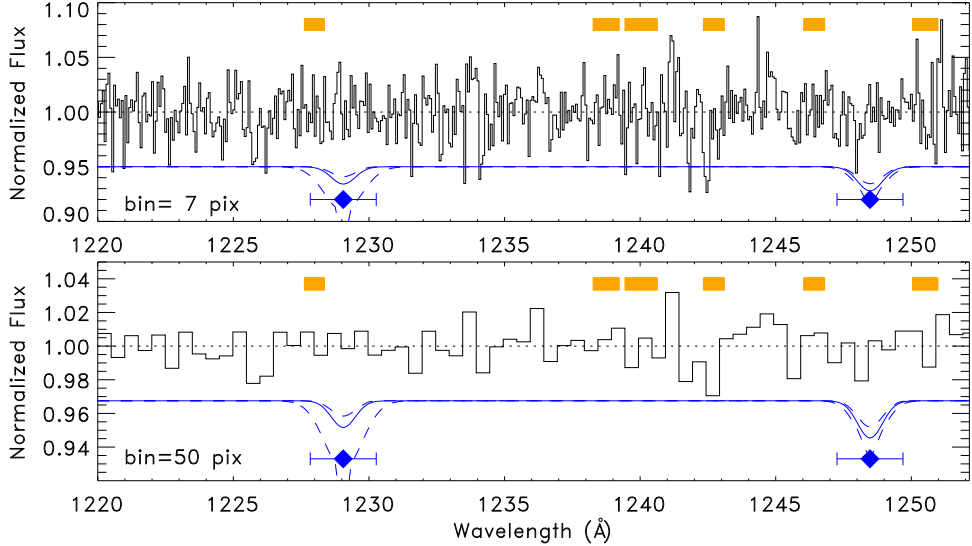} 
  \caption{Predicted broad \Lya\ absorbers superimposed on the
  normalized, line-less COS spectrum of Mrk\,421.  In the top panel,
  the COS data are binned to seven pixels ($\sim1$ resolution element,
  $\sim17$ \kms).  Orange bars show regions where narrow absorption
  features have been modeled out of the spectrum.  The blue curves
  offset below the data show the range of BLA absorption profiles
  predicted from the O\,VII detections (solid blue lines) and
  $\pm$1$\sigma$ uncertainties in $T$ and $N({\rm O\,VII})$ (dashed
  blue lines) of \citet{Nicastro05a} as listed in Table~1.  Blue
  diamonds and horizontal error bars show the velocity centroids and
  $\pm300$ \kms\ uncertainty of the \Chandra\ detections.  The bottom
  panel shows the improvement in $(S/N)_x$ obtained by binning the
  data by 50 pixels.  Optimal binning for BLAs at $T\sim10^6$~K is
  $\sim100$ pixels.}
\end{figure*}

\subsection{Signal-to-Noise and Significance Levels} 

The predicted BLAs are very weak (line-center optical depth $\tau_0<3$\%) and much broader than the $\sim18$~\kms\ instrumental resolution of COS.  They will be difficult to detect in even high-quality HST/COS spectra and will require a good understanding of the signal-to-noise ratio and systematics in the data.  The per-pixel scatter in the normalized, coadded data gives signal-to-noise per pixel $(S/N)_1\equiv \sigma^{-1}=25\pm1$ over the \Lya\ forest wavelength range.  In the Poissonian ideal, pixel-to-pixel noise is uncorrelated, and combining adjacent pixels results in lower noise (and hence higher S/N) by the square root of the number of pixels binned.  We find that binning or smoothing the line-less, normalized data by $x$ pixels results in an empirical relationship $(S/N)_x=(S/N)_1\,x^{0.38\pm0.02}$, consistent with the behavior in other datasets over a wider wavelength range \citep{Keeney11}.  The fact that the index on $x$ is less than $0.5$ shows that the noise in adjacent pixels {\it is} correlated to some degree, i.e., fixed pattern noise exists in the data. 
 
The significance level of a detection depends on the observed equivalent width $W_\lambda$ and the S/N of the data.  Following the method of \citet{Keeney11} (their equation A6), we calculate the significance level, $SL$, of an absorption feature in data binned by $x$ pixels as 
\begin{equation}
	SL(x)=\frac{f_c(x)\,W_\lambda\,(S/N)_x}{\Delta\lambda_x}=\frac{f_c(x)\,W_\lambda\,(S/N)_1\,x^{0.38}}{x\,\Delta\lambda_0},
\end{equation} 
where $f_c(x)$ is the fraction of the total line profile encompassed by $x$ pixels.  For lines much broader than the instrumental resolution, \citeauthor{Keeney11} find that $SL(x)$ is maximized when the data are binned to approximately the FWHM of the line or $x\approx2\sqrt{\ln 2}\,b$ and that this binning encompasses $f_c(x)=76\%$ of the total line area.  Since each COS pixel in the region of the \Lya\ forest is approximately 2.4 \kms, this means
\begin{equation}
  	SL(b)=0.095\times W_\lambda\,\frac{(S/N)_1}{b^{0.62}}\;,\label{eq:slb}
\end{equation}
with $W_\lambda$ in units of m\AA\ and $b$ in \kms.  

Applying Eq.~\ref{eq:slb} to observed lines, we confirm that the \Lya\ forest line at 1228.0 \AA\ is highly significant ($SL\sim30\sigma$).  The weak absorption feature at 1246.4 \AA\ assumed to be \Lya\ at $z=0.0253$ is a $\sim5\sigma$ detection in optimally-binned data.  Similarly, the stronger (1238.82~\AA) line of the Galactic \NV\ doublet is a $\sim9\sigma$ detection and the blended Galactic \ion{Mg}{2} doublet is significant at $\sim3\sigma$.

\subsection{WHIM Parameter Space Accessible by BLA}

Figure~2 shows the temperature and total column density parameter space in which we might expect WHIM systems.  HST/COS data with $(S/N)_1=25$ are sensitive to BLAs in the upper left half of the parameter space.  Also shown are the regions in which WHIM could be detected via X-ray (\OVII) and UV (\OVI) metal-line diagnostics at various column densities in $Z=0.1\,Z_\sun$.  The metal-line sensitivity limits scale on the vertical axis as $Z^{-1}$ gas.  

Can we confirm or disprove the X-ray WHIM detections based on the COS data?  Figure~3 shows the COS data with identified absorption features removed (see Fig.~1) and then binned to 7 and 50 pixels in the upper and lower panels, respectively.  The WHIM absorbers predicted from the X-ray measurements would appear at $\lambda\sim1230$ \AA\ and $\lambda\sim1249$ \AA\ (blue diamonds), both with a velocity uncertainty of $\pm300$ \kms\ from N05.  As predicted, broad features appear more significant in the lower panel where the pixel-to-pixel noise has been reduced and the binning width is closer to the characteristic line width.  The N05 WHIM absorbers modeled via Eq.~3 are predicted to produce BLAs with doppler widths $b\approx150$ \kms\ and equivalent widths of $\sim18$ and $\sim26$ m\AA\ for the $z=0.011$ and $z=0.028$ features, respectively (blue curve in Figure~3).  From Eq.~\ref{eq:slb}, these would be significant (modulo the considerable range in predicted BLA properties; see Table~1) at the $1.9\sigma$ and $2.7\sigma$ levels in $(S/N)_1=25$ HST/COS data binned to $\sim100$ pixels.  

It is clear from Figure~3 that no $\sim2-3\sigma$ absorption features appear in the COS data consistent in either line profile or centroid with the predicted WHIM absorbers.  The claimed Great Wall WHIM detection should appear at $2-3\sigma$ significance at $1248.5\pm1.2$~\AA, but is not seen in the COS data.  The 1246.4 \AA\ feature was discussed above and was removed from the residual spectrum in Figure~3.  While relatively broad for a \Lya\ feature ($b=52\pm12$ \kms; $T\la10^5$~K), it is inconsistent with the \citeauthor{Nicastro05a} \OVII\ detection at $cz=8090\pm300$ \kms\ in both redshift and inferred temperature.  The 1228.0 \AA\ \Lya\ line is consistent in redshift with the claimed $z=0.011$ \OVII\ system, but is far too narrow to arise in gas at $T\sim10^6$~K.  

We therefore rule out WHIM absorption for the two systems with the redshift, temperature and metallicity reported by \citet{Nicastro05a} at a $\ga2\sigma$ level.  However, if lower \OVII\ column densities, higher temperatures, or higher metallicities are assumed, the ($N_H,T$) parameter space locations of the claimed WHIM systems move toward the lower right in Figure~2 and would not be detectable in the current COS data.

Two additional weak features become apparent in Figure~3 in locations unrelated to the \OVII-predicted BLAs.  While both appear significant in the highly-binned lower panel of Figure~3, they are best fitted as single, narrow components more typical of the photoionized \Lya\ forest.  The first is a feature near 1225.8 \AA, a region uncontaminated by narrow absorption lines.  It is well-fitted as a $\sim4\sigma$ narrow \Lya\ system ($b=25\pm11$ \kms, $W_\lambda=13\pm4$ m\AA, $\log\,N_{\rm HI}=12.4\pm0.1$) at $z=0.0083$.  The second residual feature appears as an absorption excess in the weaker Galactic \NV\ line at 1242.5 \AA.  Measurement of this feature is complicated by the overlying \NV\ absorption, but if the absorption profile of the stronger \NV\ line is applied to the weaker line, the residual can be fitted as a $\sim5\sigma$ feature: $b\sim27$ \kms, $W_\lambda\sim19$ m\AA, $\log\,N_{\rm HI}\sim12.6$, $z=0.0220$ if interpreted as \Lya.

\section{Conclusions and Implications}

The \citet{Nicastro05a} claims of \OVII\ absorbers toward Mrk\,421 were met with a great deal of skepticism in the X-ray community \citep{Kaastra06,Rasmussen07}.  Even the more significant $z=0.027$ \OVII\ system in the Great Wall ($N_{\rm OVII}\approx7\times10^{14}$~\cd) is only a $\sim2$~m\AA\ feature in the \Chandra\ LETG data and relies critically on the continuum model in the $21-23$~\AA\ region.  While Mrk\,421 has been observed for a total of $>200$ ksec over many different observing campaigns and features some of the highest $S/N$ data for any extragalactic X-ray target, a more reliable \OVII\ detection limit may require $N_{\rm OVII}\ga3\times10^{15}$~\cd\ ($\ga5$~m\AA).  Note that the strong ($N_{\rm OVII}\sim10^{16}$~\cd) \OVII\ absorption at $z=0$ toward Mrk\,421 is uncontroversial.  Indeed, it is typical of $z=0$ absorption seen in numerous extragalactic sight lines \citep{Fang02,Fang03,McKernan04,Nicastro02,Nicastro05a,Wang05,Williams06,Buote09} and is thought to arise in hot gas in the Galactic halo.

In this paper we highlight the diagnostic power of thermally-broadened \Lya\ absorbers in far-UV spectra as tracers of WHIM gas.  Mrk\,421 was observed for less than half an hour (1.7 ksec or half an \HST\ orbit) in the blue G130M mode of the Cosmic Origins Spectrograph, and yet it reaches WHIM detection levels comparable to those in $>200$ ksec of X-ray observations with \Chandra.  Based on these high-quality data, we can rule out the existence of WHIM gas over a large region of ($T,N_H$) parameter space (see Figures~2, 3) along much of the Mrk\,421 sight line.  Specifically, we rule out the existence of broad \Lya\ counterpart absorption to the claimed \OVII\ WHIM absorbers at $z=0.011$ and $z=0.027$ \citep{Nicastro05a,Nicastro05b} at $\sim2\sigma$ level.  We cannot, however, rule out the existence of the claimed WHIM absorbers if their metallicities are higher than the assumed $Z>0.1\,Z_\sun$, if the \OVII\ column densities are at the lower limits of their reported ranges, or if the temperatures are higher than the $\sim1-2\times10^6$~K inferred from ion ratios \citep{Nicastro05a,Nicastro05b,Savage05}.  The \Lya\ absorption in any of these systems would then be below the detection limit of the current COS data.

The search for missing baryons has long centered on the gas in the WHIM phase via one of several techniques.  Roughly one hundred low-$z$ \OVI\ detections have been reported in $\sim30$ sight-lines \citep{DS08,Tripp08,ThomChen08} which give a statistically-significant sample of lower temperature, higher metallicity WHIM baryons ($T<10^6$~K; $Z>0.1\,Z_\sun$; see red contours in Figure~2).  However, while the presence of these absorbers is agreed upon in most cases, their interpretation as WHIM gas is still controversial in roughly half the cases \citep[see,][]{Danforth09}.  X-ray metal lines such as \OVII\ seem to present a more secure identification of WHIM absorption, but suffer from small-number statistics and are still at or beyond the cutting edge of current X-ray instruments.  Different, complementary techniques must be employed to verify the phase membership of the \OVI\ systems and confirm the presence of X-ray WHIM absorbers.  

The most convincing X-ray WHIM detections have relied on targeted searches of nearby supercluster filaments like the Sculptor Wall detection \citep{Buote09,Fang10}, the Pisces-Cetus Supercluster, and the Far Sculptor Wall \citep{Zappacosta10}.  The Sculptor Wall (at $z=0.030\pm0.002$ in the direction of the blazar target H\,2356$-$309) used both \Chandra\ ACIS and \XMM\ RGS to detect the \OVII\ K$\alpha$ absorption line at $z=0.032$.  The \OVIII\ detection at $z=0.055$ \citep{Fang02} towards PKS\,2155$-$304 is coincident with a group of narrow \Lya\ lines at the redshift of a group of spiral galaxies \citep{Shull98,Shull03}.  In many ways, the search for WHIM along the Mrk\,421 sight line (whether via X-ray or UV metal ions or BLAs) should also be considered a targeted search, since this object lies just behind a section of the Great Wall, whether or not WHIM gas is ultimately confirmed.

Although the current dataset appears to rule out the WHIM detections claimed by N05, due to the imprtance of these claims, we plan to re-observe Mrk\,421 with COS in \HST\ Cycle 19.  Deeper observations will either reveal a BLA feature below the current detection limit or place stronger constraints on WHIM gas in the sight line.  Either way, the detection/non-detection of WHIM in the Great Wall will add an important data-point to the correlation of hot baryons and large-scale structure.  The covering factor of WHIM gas around galaxy filamentary structure has significant impact on our understanding of the missing baryon problem at low-$z$.  Correlating the much larger catalogs of low-$z$ \OVI\ detections \citep{DS08,Tripp08,ThomChen08} with galaxy redshift surveys may reveal additional filament-WHIM correspondence, and document the WHIM non-detections in similar LSS features.

\medskip

This work was supported by NASA grants NNX08AC146 and NAS5-98043 to the University of Colorado at Boulder.  CWD acknowledges feedback from F. Nicastro and M. Elvis.

{\it Facilities}: HST(COS)

\end{document}